\documentclass[seceq]{ptptex}

\usepackage{graphicx}

\title{
Light Scalar Mesons in the QCD Sum Rule }

\author{
Hua-Xing \textsc{Chen}$^{1,2,}$\footnote{ e-mail address:
hxchen@rcnp.osaka-i.ac.jp}, Atsushi \textsc{Hosaka}$^{1,}$\footnote{
e-mail address: hosaka@rcnp.osaka-u.ac.jp} and Shi-Lin
\textsc{Zhu}$^{2,}$\footnote{ e-mail address:
zhusl@th.phy.pku.edu.cn} }

\inst{
$^1$Research Center for Nuclear Physics, Osaka University, Ibaraki
567--0047, Japan \\$^2$Department of Physics, Peking University,
Beijing 100871, China }

\abst{
We study the light scalar mesons in the QCD sum rule. We find that
there are five independent scalar tetraquark currents in the local
form, and we perform QCD sum rule analysis using both these currents
and their combinations. Compared with the sum rules by conventional
$\bar q q$ currents, our result supports a tetraquark structure for
low-lying scalar mesons. }

\begin{document}

\maketitle

\section{Introduction}
The nature of light scalar mesons of $up$, $down$ and $strange$
quarks is not fully understood~\cite{scalar,Yao:2006px}. The
expected members are $\sigma(600)$, $\kappa(800)$, $f_0(980)$ and
$a_0(980)$ forming a nonet of flavor SU(3). Because they have the
same spin and parity as the vacuum, $J^P = 0^+$, they reflect the
bulk properties of the non-perturbative QCD vacuum. So far, several
different pictures for the scalar mesons have been proposed. In the
conventional quark model, they have a $\bar q q$ configuration of
$^3P_0$ whose masses are expected to be larger than 1 GeV due to the
$p$-wave orbital excitation. Furthermore, the mass ordering in a
naive quark mass counting of $m_u \sim m_d < m_s$ implies $m_\sigma
\sim m_{a_0} < m_\kappa < m_{f_0}$. In chiral models, they are
regarded as chiral partners of the Nambu-Goldstone bosons ($\pi, K,
\eta, \eta^\prime)$~\cite{Hatsuda:1994pi}. Due to the collective
nature, their masses are expected to be lower than those of the
quark model. Yet another interesting picture is that they are
tetraquark
states~\cite{Jaffe:1976ig,Lee:2006vk,Brito:2004tv,Zhang:2006xp}. In
contrast with the $\bar q q$ states, their masses are expected to be
around 0.6 -- 1 GeV with the ordering  of $m_\sigma <  m_\kappa <
m_{f_0, a_0}$, consistent with the recent experimental observation
~\cite{scalar,Yao:2006px,experiment}. If such tetraquarks survive,
they may be added to members of exotic multiquark states.

In this contribution, we would like to report the results of a
systematic study of the masses of the tetraquark scalar mesons in
the QCD sum rule. We find that the QCD sum rule analysis with
tetraquark currents implies the masses of scalar mesons in the
region of 600 -- 1000 MeV with the ordering, $m_\sigma < m_\kappa <
m_{f_0, a_0}$, while the conventional $\bar q q$ currents imply
masses around 1.5 GeV.

\section{Independent Currents}

Let us start with currents for the scalar tetraquark, which we
consider only local currents. Using the antisymmetric combination
for diquark flavor structure, we arrive at the following five
independent currents~\cite{Chen:2006hy}
\begin{eqnarray}
\nonumber\label{define_udud_current} S^\sigma_3 &=& (u_a^T C
\gamma_5 d_b)(\bar{u}_a \gamma_5 C \bar{d}_b^T - \bar{u}_b \gamma_5
C \bar{d}_a^T)\, ,
\\ \nonumber
V^\sigma_3 &=& (u_a^T C \gamma_{\mu} \gamma_5 d_b)(\bar{u}_a
\gamma^{\mu}\gamma_5 C \bar{d}_b^T - \bar{u}_b \gamma^{\mu}\gamma_5
C \bar{d}_a^T)\, ,
\\
T^\sigma_6 &=& (u_a^T C \sigma_{\mu\nu} d_b)(\bar{u}_a
\sigma^{\mu\nu} C \bar{d}_b^T + \bar{u}_b \sigma^{\mu\nu} C
\bar{d}_a^T)\, ,
\\ \nonumber
A^\sigma_6 &=& (u_a^T C \gamma_{\mu} d_b)(\bar{u}_a \gamma^{\mu} C
\bar{d}_b^T + \bar{u}_b \gamma^{\mu} C \bar{d}_a^T)\, ,
\\ \nonumber
P^\sigma_3 &=& (u_a^T C d_b)(\bar{u}_a C \bar{d}_b^T - \bar{u}_b C
\bar{d}_a^T)\, ,
\end{eqnarray}
where the sum over repeated indices ($\mu$, $\nu, \cdots$ for Dirac,
and $a, b, \cdots$ for color indices) is taken. Either plus or minus
sign in the second parentheses ensures that the diquarks form the
antisymmetric combination in the flavor space. The currents $S$,
$V$, $T$, $A$ and $P$ are constructed by scalar, vector, tensor,
axial-vector, pseudoscalar diquark and antidiquark fields,
respectively. The subscripts $3$ and $6$ show that the diquarks
(antidiquark) are combined into the color representation
$\mathbf{\bar 3_c}$ and $\mathbf{6_c}$ ($\mathbf{3_c}$ or
$\mathbf{\bar 6_c}$), respectively. The currents for other members
are formed similarly. We can also use a symmetric combination for
diquark flavor structure. However, they are related to the
antisymmetric ones by the axial U(1)
transformation~\cite{Umekawa:2004js}.

\section{QCD Sum Rule Analysis}
For the past decades QCD sum rule has proven to be a very powerful
and successful non-perturbative
method~\cite{Shifman:1978bx,Reinders:1984sr}. In sum rule analyses,
we consider two-point correlation functions:
%
\begin{equation}
\Pi(q^2)\,\equiv\,i\int d^4x e^{iqx}
\langle0|T\eta(x){\eta^\dagger}(0)|0\rangle \, , \label{eq_pidefine}
\end{equation}
%
where $\eta$ is an interpolating current for the tetraquark. We
compute $\Pi(q^2)$ in the operator product expansion (OPE) of QCD up
to certain order in the expansion, which is then matched with a
hadronic parametrization to extract information of hadron
properties. At the hadron level, we express the correlation function
in the form of the dispersion relation with a spectral function:
%
\begin{equation}
\Pi(p)=\int^\infty_0\frac{\rho(s)}{s-p^2-i\varepsilon}ds \, ,
\label{eq_disper}
\end{equation}
%
where
%
\begin{eqnarray}
\rho(s) & \equiv & \sum_n\delta(s-M^2_n)\langle
0|\eta|n\rangle\langle n|{\eta^\dagger}|0\rangle \ \nonumber\\ &=&
f^2_X\delta(s-M^2_X)+ \rm{higher\,\,states}\, . \label{eq_rho}
\end{eqnarray}
%
For the second equation, as usual, we adopt a parametrization of one
pole dominance for the ground state $X$ and a continuum
contribution. The mass of the state $X$ can be obtained
%
\begin{equation}
M^2_X=\frac{\int^{s_0}_0 e^{-s/M_B^2}s\rho(s)ds}{\int^{s_0}_0
e^{-s/M_B^2}\rho(s)ds}\, . \label{eq_LSR}
\end{equation}
%

We performed the sum rule analysis using all currents and their
various linear combinations, and found a good sum rule by a linear
combination of $A_6^\sigma$ and $V_3^\sigma$
\begin{eqnarray}
\eta^\sigma_1 = \cos\theta A^\sigma_6 + \sin\theta V^\sigma_3\, ,
\end{eqnarray}
where the best choice of the mixing angle turns out to be
$\cot\theta = 1 / \sqrt{2}$. For $\kappa$, $f_0$ and $a_0$, we have
also found that similar linear combinations give better sum rules.
The results of OPE can be found in Ref.~\cite{Chen:2006zh}

\section{Numerical Analysis}

For numerical calculations, we use the following values of
condensates~\cite{Yang:1993bp,Ioffe:2002be,Gimenez:2005nt}:
$\langle\bar qq \rangle=-(0.240 \mbox{ GeV})^3$, $\langle\bar
ss\rangle=-(0.8\pm 0.1)\times(0.240 \mbox{ GeV})^3$,$\langle
g_s^2GG\rangle =(0.48\pm 0.14) \mbox{ GeV}^4$, $ m_u = 5.3 \mbox{
MeV}$, $m_d = 9.4 \mbox{ MeV}$, $m_s(1\mbox{ GeV})=125 \pm 20 \mbox{
MeV}$, $\langle g_s\bar q\sigma G q\rangle=-M_0^2\times\langle\bar
qq\rangle$, $M_0^2=(0.8\pm0.2)\mbox{ GeV}^2$.

\begin{figure}[hbt]
\begin{center}
\scalebox{0.8} {\includegraphics{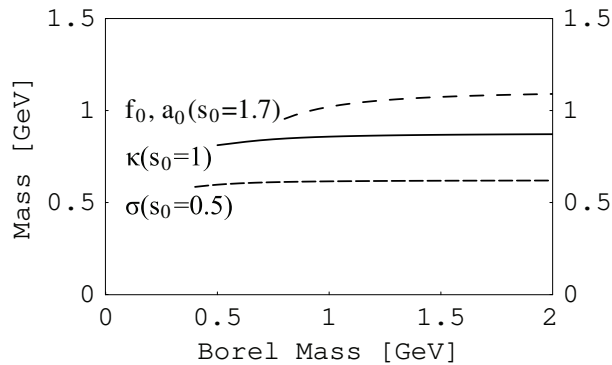}} \caption{Masses of
the $\sigma$ (short-dashed), $\kappa$ (solid), $f_0$ and $a_0$
(long-dashed) mesons calculated by the tetraquark currents as
functions of the Borel mass $M_B$, with $s_0$ (GeV$^2$) as shown in
figures.} \label{pic_tetra}
\end{center}
\end{figure}

The sum rules are written as power series of the Borel mass $M_B$.
Since the Borel transformation suppresses the contributions from $s
> M_B$, smaller values are preferred to suppress the continuum
contributions also. However, for smaller $M_B$ convergence of the
OPE becomes worse. Therefore, we should find an optimal $M_B$
preferably in a small value region. We have found that the minima of
such a region are 0.4 GeV for $\sigma$, 0.5 GeV for $\kappa$ and 0.8
GeV for $f_0$ and $a_0$, where the pole contributions reach around
50 \% for all cases~\cite{Chen:2006zh}. As $M_B$ is increased, the
pole contributions decrease, but the resulting tetraquark masses are
stable as shown in Fig.~\ref{pic_tetra}.

After careful test of the sum rule for a wide range of parameter
values of $M_B$ and $s_0$, we have found reliable sum rules, with
which we find the masses $ m_\sigma = (0.6 \pm 0.1) \; {\rm GeV}$, $
m_\kappa = (0.8 \pm 0.1) \; {\rm GeV}$,  $m_{f_0,a_0} = (1 \pm 0.1)
\; {\rm GeV}\; ,$ which are consistent with the experimental
results~\cite{Yao:2006px}.

For comparison, we have also performed the QCD sum rule analysis
using the $\bar q q$ current within the present framework. The
stable (weak $M_B$) behavior is obtained with the masses of all four
mesons around 1.5 GeV. Here again we have tested various values of
$M_B$ and $s_0$, and confirmed that the result shown is optimal.

\section{Conclusions}

We have performed the QCD sum rule analysis with tetraquark
currents, which implies the masses of scalar mesons in the region of
600 -- 1000 MeV with the ordering, $m_\sigma < m_\kappa < m_{f_0,
a_0}$. We have also performed the QCD sum rule analysis with the
conventional $\bar q q$ currents, which implies masses around 1.5
GeV. We have tested all possible independent tetraquark currents as
well as their linear combinations. Our observation supports a
tetraquark structure for low-lying scalar mesons. To test the
validity of the tetraquark structure, it is also important to study
decay properties, which is often sensitive to the structure of wave
functions. Such a tetraquark structure will open an alternative path
toward the understanding exotic multiquark dynamics which one does
not experience in the conventional hadrons.

\section*{Acknowledgements}
H.~X.~C and A.~H. thank the Yukawa Institute for Theoretical Physics
at Kyoto University for hospitality during the YKIS2006 on "New Frontiers on QCD".
H.X.C. is grateful to the Monkasho fellowship for supporting his
stay at RCNP, Osaka University. A.H. is supported in part by the
Grant for Scientific Research ((C) No.16540252) from the Ministry of
Education, Culture, Science and Technology, Japan. S.L.Z. was
supported by the National Natural Science Foundation of China under
Grants 10375003 and 10421503, Ministry of Education of China,
FANEDD, Key Grant Project of Chinese Ministry of Education (NO
305001) and SRF for ROCS, SEM.


\begin{thebibliography}{10}

\bibitem{scalar}
  E. M. Aitala et al., Phys. Rev. Lett. 86, 770 (2001); M. Ablikim
  et al., Phys. Lett. B 598, 149 (2004).

\bibitem{Yao:2006px}
  W.~M.~Yao {\it et al.}  [Particle Data Group],
  J.\ Phys.\ G {\bf 33}, 1 (2006).

\bibitem{Hatsuda:1994pi}
  T.~Hatsuda and T.~Kunihiro,
  Phys.\ Rept.\  {\bf 247}, 221 (1994).

\bibitem{Jaffe:1976ig}
  R.~L.~Jaffe,
  Phys.\ Rev.\ D {\bf 15}, 267 (1977).

\bibitem{Lee:2006vk}
  H.~J.~Lee and N.~I.~Kochelev,
  Phys.\ Lett.\  B {\bf 642}, 358 (2006).

\bibitem{Brito:2004tv}
  T.~V.~Brito, F.~S.~Navarra, M.~Nielsen and M.~E.~Bracco,
  Phys.\ Lett.\ B {\bf 608}, 69 (2005).

\bibitem{Zhang:2006xp}
  A.~Zhang, T.~Huang and T.~G.~Steele,
  arXiv:hep-ph/0612146.

\bibitem{experiment}
  E. M. Aitala et al., Phys. Rev. Lett. 89, 121801 (2002); M. Ablikim
  et al., Phys. Lett. B 633, 681 (2006).

\bibitem{Chen:2006hy}
  H.~X.~Chen, A.~Hosaka and S.~L.~Zhu,
  Phys. Rev. D 74, 054001 (2006).

\bibitem{Umekawa:2004js}
  T.~Umekawa, K.~Naito, M.~Oka and M.~Takizawa,
  Phys.\ Rev.\  C {\bf 70}, 055205 (2004).


\bibitem{Shifman:1978bx}
  M.~A.~Shifman, A.~I.~Vainshtein and V.~I.~Zakharov,
  Nucl.\ Phys.\ B {\bf 147}, 385 (1979).

\bibitem{Reinders:1984sr}
  L.~J.~Reinders, H.~Rubinstein and S.~Yazaki,
  Phys.\ Rept.\  {\bf 127}, 1 (1985).

\bibitem{Yang:1993bp}
  K.~C.~Yang, W.~Y.~P.~Hwang, E.~M.~Henley and L.~S.~Kisslinger,
  Phys.\ Rev.\ D {\bf 47}, 3001 (1993).

\bibitem{Ioffe:2002be}
  B.~L.~Ioffe and K.~N.~Zyablyuk,
  Eur.\ Phys.\ J.\ C {\bf 27}, 229 (2003).

\bibitem{Gimenez:2005nt}
  V.~Gimenez, V.~Lubicz, F.~Mescia, V.~Porretti and J.~Reyes,
  Eur.\ Phys.\ J.\ C {\bf 41}, 535 (2005).

\bibitem{Chen:2006zh}
  H.~X.~Chen, A.~Hosaka and S.~L.~Zhu,
  arXiv:hep-ph/0609163.

\end{thebibliography}
\end{document}